\documentclass[12pt]{article}

\usepackage{psfig}

\usepackage{scicite}

\usepackage{times}

\newenvironment{sciabstract}{%
\begin{quote} \bf}
{\end{quote}}

\newcounter{lastnote}
\newenvironment{scilastnote}{%
\setcounter{lastnote}{\value{enumiv}}%
\addtocounter{lastnote}{+1}%
\begin{list}%
{\arabic{lastnote}.}
{\setlength{\leftmargin}{.22in}}
{\setlength{\labelsep}{.5em}}}
{\end{list}}

\title{A Double-Pulsar System --- A Rare Laboratory for
Relativistic Gravity and Plasma Physics}

\author{A.G. Lyne,$^{1\ast}$ M. Burgay,$^{2}$ M. Kramer,$^{1}$ A. Possenti,$^{3,4}$ \\ 
R.N. Manchester,$^{5}$ F. Camilo,$^{6}$ M.A. McLaughlin,$^{1}$ D.R. Lorimer,$^{1}$ \\
N. D'Amico,$^{3,7}$ B.C. Joshi,$^{8}$ J. Reynolds$^{9}$ and P.C.C. Freire$^{10}$ \\
\\
\normalsize{$^{1}$University of Manchester, Jodrell Bank Observatory, Macclesfield, SK11 9DL, UK}\\
\normalsize{$^{2}$Universit\`a degli Studi di Bologna, Dipartimento di Astronomia, via Ranzani 1,}\\
\normalsize{40127 Bologna, Italy}\\
\normalsize{$^{3}$INAF - Osservatorio Astronomico di Cagliari, Loc. Poggio dei Pini, Strada 54,}\\
\normalsize{09012 Capoterra, Italy}\\
\normalsize{$^{4}$INAF - Osservatorio Astronomico di Bologna, via Ranzani 1, 40127 Bologna, Italy}\\
\normalsize{$^{5}$Australia Telescope National Facility, CSIRO, P.O.~Box~76, Epping,
NSW~1710, Australia}\\
\normalsize{$^{6}$Columbia Astrophysics Laboratory, Columbia University, 550 West 120th Street,}\\
\normalsize{New York, NY 10027, USA}\\
\normalsize{$^{7}$Universit\`a degli Studi di Cagliari, Dipartimento di Fisica, SP Monserrato-Sestu km 0.7,}\\
\normalsize{09042 Monserrato, Italy}\\
\normalsize{$^{8}$National Centre for Astrophysics, P.O. Bag 3, Ganeshkhind, Pune 411007, India}\\
\normalsize{$^{9}$ATNF, Parkes Observatory, P.O. Box 276, Parkes, NSW 2870, Australia}\\
\normalsize{$^{10}$NAIC, Arecibo Observatory, HC03 Box 53995, PR 00612, USA}\\
\\
\normalsize{$^\ast$To whom correspondence should be addressed; E-mail: agl@jb.man.ac.uk}
}

\date{}

\begin{document} 


\baselineskip24pt


\maketitle


\begin{sciabstract}
The clock-like properties of pulsars moving in the gravitational
fields of their unseen neutron-star companions have allowed unique
tests of general relativity and provided evidence for gravitational
radiation.  We report here the detection of the 2.8-sec pulsar
J0737$-$3039B as the companion to the 23-ms pulsar J0737$-$3039A in a
highly-relativistic double-neutron-star system, allowing unprecedented
tests of fundamental gravitational physics. We observe a short eclipse
of J0737$-$3039A by J0737$-$3039B and orbital modulation of the flux
density and pulse shape of J0737$-$3039B, probably due to the
influence of J0737$-$3039A's energy flux upon its magnetosphere.
These effects will allow us to probe magneto-ionic properties of a
pulsar magnetosphere.

\end{sciabstract}


Double-neutron-star (DNS) binaries are rare, and only six such systems
are known.  However, they can provide wonderful laboratories for the
study of relativistic gravity and gravitational radiation
\cite{tay92}.  The discovery of such systems has been a prime
objective of pulsar surveys since the first, PSR~B1913+16,
was discovered 29 years ago \cite{ht75a}.  Apart from their rarity,
they are particularly difficult to find because of the large orbital
acceleration experienced by the members of the systems, resulting in
large and varying Doppler effects in the observed rotational period.

The pulsar J0737$-$3039 was recently discovered as part of a
high-latitude multibeam survey of the southern sky using the 64-m
Parkes radio telescope \cite{bdp+03}.  It was found to be in a 2.4-hr
eccentric orbit with another compact object that the observed orbital
parameters suggested was another neutron star.  The short orbital
period and compactness of the system and the high timing precision
made possible by the large flux density and narrow pulse features of
this pulsar promise to make this system a superb laboratory for the
investigation of relativistic astrophysics.  Relativistic advance of
the angle of periastron has already been measured and orbital decay
due to gravitational wave emission will be measured with high
precision within a few months.  The resulting in-spiral will end in
coalescence of the two stars in about 85~My \cite{bdp+03}.  This discovery
significantly increases the estimates of the detection rate of DNS
in-spirals by gravitational wave detectors \cite{bdp+03,kkl+04}.

\paragraph*{The discovery of PSR~J0737$-$3039B.}

No other pulsar was detected in the 4.5-min discovery observation of
PSR~J0737$-$3039.  However, analysis of data acquired subsequently for
detailed studies of the system has revealed the occasional presence of
pulsations with a period of 2.8 seconds.  This pulsar, henceforth
called PSR~J0737$-$3039B (or simply B in this paper), has the same
dispersion measure (the integrated column density of free electrons
along the line of sight) as the original pulsar, henceforth called
PSR~J0737$-$3039A (or simply A) and shows Doppler variations in the
pulse period which identify it as the companion.

The data obtained by Burgay et al. \cite{bdp+03} for their studies of
A used the Parkes radio telescope at a frequency of 1390~MHz and have
now all been reprocessed to enable investigation of the properties of
B.  Additionally, the system was observed on seven days in November
2003 simultaneously with dual-polarization receivers centered on
frequencies of 680~MHz and 3030~MHz \cite{bandwidth}.  Observations
were also made of both pulsars at 1396~MHz with the 76-m Lovell
telescope at Jodrell Bank, also with a dual-polarization receiver.
Timing observations of both pulsars have been made covering the 7
months from May to November 2003, with individual data spans between
10 minutes and 5 hours.

The main features of the observations of B are shown in Fig.~1, which
illustrates the strength of the 2.8-sec pulsed emission in the three
frequency bands as a function of orbital phase and pulse phase over
the full orbit.  Most notable is the variation in the received flux
density of the pulsar, that is clearly visible for two brief periods
of about 10 minutes duration each centered upon orbital longitudes
210$^\circ$ and 280$^\circ$.  Within these bursts, the strength is
such that most pulses are detected individually.  In addition to the
orbital phases where it is always detected, the pulsar often shows
weak emission elsewhere, notably between orbital phases 0$^\circ$ and
20$^\circ$ (Fig.~1B).  It is likely that observations with greater
sensitivity will reveal more widespread emission throughout the orbit.
The pattern of the visibility is essentially stable from
orbit to orbit and from band to band over the full frequency range of
680~MHz to 3030~MHz.  The absence of pulsar B in the discovery
observation of A can now be understood, as that observation was made
at a longitude of 146$^\circ$, where there is little emission from B.

There are significant changes in the shape of the pulses of B with
orbital phase (Fig.~1).  The data at 1390~MHz show that it varies from
a narrow intense main pulse with a weak precursor in the burst around
longitude 210$^\circ$, to a roughly-equal double-component profile
with a somewhat greater separation near longitude 280$^\circ$. It
becomes essentially a single pulse at around longitude 0$^\circ$.
There is also an indication for frequency evolution of
pulse shape within the bursts.  The double nature of the profiles
becomes more evident at the higher frequency of 3030~MHz and less so
at 680~MHz, where the components seem to have nearly merged.

In order to make timing measurements of the pulsar, a phase-dependent
set of pulse templates was created by forming average profiles for
each of the three burst regions.  These templates were aligned in such
a way that the trailing, usually more intense, components were at the
same pulse phase.  Cross-correlation of data with appropriate
templates from this set was carried out to produce times of arrival
(TOAs).  These TOAs were then compared with a model for the
astrometric, rotational and orbital parameters using the TEMPO pulsar
timing program \cite{tempo}.  The position and main orbital parameters
have been determined from the high-precision timing enabled by the
strong, narrow pulses of A (Table~1). For B, an offset of 180$^\circ$
was added to the longitude of periastron and the only fitted
parameters were the pulsar rotational period, $P$, its first
derivative, $\dot{P}$, and the projected semi-major axis,
$x_{\rm{B}}=a_{\rm{B}} \sin i/c$, where $i$ is the orbital inclination
and $c$ is the speed of light (Table~1).

\paragraph*{Gravitational Physics.}

With their strong gravitational fields and rapid motions, DNS binaries
exhibit large relativistic effects.  General relativity and other
theories of gravity can be tested when a number of relativistic
corrections, the so-called post-Keplerian (hereafter PK) parameters,
to the classical Keplerian description can be measured. In this
formalism, for point masses with negligible spin contributions, the PK
parameters in each theory should only be functions of the a priori
unknown neutron star masses and the classical Keplerian
parameters. With the two masses as the only free parameters, the
measurement of three or more PK parameters over-constrains the system,
and thereby provides a test-ground for theories of gravity
\cite{dt92}.  In a theory that describes the binary system correctly,
the PK parameters define lines in a mass-mass diagram that all
intersect in a single point. Such tests have been possible to date in
only two DNS systems, PSR~B1913+16 \cite{tw89} and PSR~B1534+12
\cite{sttw02} (see also \cite{bokh03}).  For PSR~B1913+16, the
relativistic periastron advance, $\dot{\omega}$, the orbital decay due
to gravitational wave damping, $\dot{P}_{\rm b}$, and the
gravitational redshift/time dilation parameter, $\gamma$, have been
measured, providing a total of three PK parameters.  For PSR~B1534+12,
Shapiro delay \cite{dd86}, caused by passage of the pulses through the
gravitational potential of the companion, is also visible, since the
orbit is seen nearly edge-on. This results in two further PK
parameters, $r$ (``range'') and $s$ (``shape'') of the Shapiro delay.
However, the observed value of $\dot{P}_{\rm b}$ requires correction
for kinematic effects \cite{dt91}, so that PSR~B1534+12 provides four
PK parameters usable for precise tests \cite{sttw02}.

Extending and improving the timing solution for A \cite{bdp+03}, we
have measured A's $\dot{\omega}$ and $\gamma$ and have also detected
the Shapiro delay in the pulse arrival times of A due to the
gravitational field of B (Fig.~2). This provides four measured PK
parameters, resulting in a $m_A-m_B$ plot (Fig.~3) through which we
can test the predictions of general relativity
\cite{dd85,dd86}. However, the detection of B as a pulsar opens up
opportunities to go beyond what is possible with previously known
DNS binary systems. Firstly, we can exclude all regions in the
$m_{\rm{A}}-m_{\rm{B}}$ plane that are forbidden by the individual
mass functions of A and B due to the requirement $\sin i\le 1$.
Secondly, with a measurement of the projected semi-major axes of the
orbits of A and B (Table~1), we obtain a precise measurement of the
mass ratio, $R(m_{\rm{A}}, m_{\rm{B}}) \equiv m_{\rm{A}}/m_{\rm{B}} =
x_{\rm{B}}/x_{\rm{A}}$, providing a further constraint in the
$m_A-m_B$ plot (Fig.~3).  This relation is valid for any theory of
gravity \cite{dt92,rhealth}.  Most importantly, the $R$-line is
independent of strong-field (self-field) effects, providing a
stringent and new constraint for tests of gravitational theories
\cite{dd85,dd86,massdef}.

With four PK parameters already available for tests, this
additional constraint makes this system the most over-determined DNS
binary to date and a truly unique laboratory for relativistic gravity.
Moreover, with a significant measurement of $\dot{P}_{\rm b}$ expected
within the next few months, an additional PK parameter will become
available. This may provide a sixth constraint if kinematic effects
are negligible or can be isolated by proper motion and distance
measurements. The position of the allowed region in Fig.~3 also
determines the inclination of the orbit to the line-of-sight.  It
turns out that the system is observed nearly edge-on with an
inclination angle $i$ of about 87$^\circ$ (Table~1).

Due to the curvature of space-time near massive objects, the spin axes
of both pulsars will precess about the total angular momentum vector,
changing the orientation of the pulsars as seen from Earth
\cite{dr74}. With the measured system parameters (Table~1), general
relativity predicts periods of such geodetic precession of only 75~yr
for A and 71~yr for B \cite{bo75}.  Hence, the relative orientation of
the pulsars' spin axes within the system geometry is expected to
change on short time-scales. This should lead to measurable changes in
the profiles of A and B (cf. \cite{wrt89,wt02,kra98}), and perhaps also to
measurable changes in the aberration effects due to the rotation of A
and B. Hence, we can expect that additional PK parameters can be
measured that are too small to be detected in other binary pulsars. One
example of these are the aberration terms $A$, $B$ and $\delta_r$ of the
DD timing model that we used \cite{dd85,dd86}, assumed to be zero for the 
present analysis.

In contrast to previous tests of general relativity, we may soon have
to use higher order terms than $(v/c)^2$ to describe the system
accurately, the first time this has been necessary for binary pulsar
systems.  The future high precision of the measurement of
$\dot{\omega}$ may demand this first for comparison of the observed
value of this parameter with theories of gravity \cite{ds88}.
Deviations from the value predicted by general relativity may be
caused by contributions from spin-orbit coupling \cite{bo75}, which is
about an order of magnitude larger than for PSR~B1913+16. This
potentially will allow us to measure the moment of inertia of a
neutron star for the first time \cite{ds88,wex95}.

\paragraph*{Origin and evolution of the double-pulsar system.}
The existence of DNS binaries can be understood by a binary evolution
scenario which starts with two main sequence stars (e.g. \cite{bv91}). The
initially more massive star evolves first and eventually
explodes in a supernova to form a neutron star. Under favorable
conditions, this neutron star remains bound to its companion and spins
down as a normal pulsar for the next $10^{6}$--$10^{7}$ yr. At some later
time, the remaining (secondary) star comes to the end of its
main-sequence lifetime and begins a red giant phase. Depending on the
orbital parameters of the system, the strong gravitational field of
the neutron star attracts matter from the red giant, forming an
accretion disk and making the system visible as an X-ray binary.  The
accretion of matter transfers orbital angular momentum to the neutron
star, spinning it up to short periods and dramatically reducing its
magnetic field \cite{bk74,smsn89}. A limiting spin period is reached
due to equilibrium between the magnetic pressure of the accreting
neutron star and the ram pressure of the infalling matter \cite{bv91,acw99}. 

A crucial phase in the evolution of close DNS binaries like
J0737$-$3039 is the dramatic reduction in orbital separation that
occurs when matter from the secondary star is expelled from the
system, resulting in a very compact system consisting of a helium star
and a neutron star \cite{vd73,fv75}.  A sufficiently massive helium
star will ultimately undergo a supernova explosion forming a young,
second neutron star. If the stars remain bound following this
explosion, the resulting system is a pair of neutron stars in an
eccentric orbit with very
different magnetic field strengths and hence spin-down properties, as
in fact observed here.  With masses of $m_{\rm{A}}=1.34$\,M$_{\odot}$
and $m_{\rm{B}}=1.25$\,M$_{\odot}$, A is typical of other neutron
stars with measured masses \cite{tc99,bokh03}, while B has a
significantly smaller mass than any other.

The time since the second supernova explosion can be estimated by
comparing our measurements of $P$ and $\dot{P}$ for A and B, which can
be used to compute their characteristic ages $\tau=P/(2\dot{P})$.  If
characteristic ages are good indicators of pulsars' true ages, we
expect $\tau_{\rm{A}} = \tau_{\rm{B}}$, but the observed values are
$\tau_{\rm{A}} \simeq 4 \tau_{\rm{B}}$. This discrepancy can be
reconciled by questioning one or more of the assumptions inherent in
the use of characteristic ages as estimates of true ages: a negligible
birth spin period and a non-decaying magnetic dipole braking
torque. At the very least, the post-accretion spin period of A cannot
have been negligible due to details of the accretion process discussed
above.  Simple models assuming constant magnetic dipole spin-down
\cite{acw99} predict a post-accretion spin period for A in the range
10--18 ms, enough to explain the observed discrepancy in the
characteristic ages.
 
\paragraph*{Probing pulsar magnetospheres.} 

The separation of the two pulsars in their orbits (Fig.~4 top) is
typically $\sim$900,000~km or 3 lt-s (the distance light travels in
three seconds).  The large orbital inclination means that, at
conjunction, the line-of-sight to one pulsar passes within about
0.15~lt-s of the other (Fig.~4 bottom).  This is substantially smaller
than the 0.45~lt-s radius of B's light cylinder, the point at which the
co-rotation speed equals the speed of light, although much
greater than the 0.004~lt-s light cylinder radius of A.  As the
pulsars move in their orbits, the line-of-sight from A passes through,
and sweeps across, the magnetosphere of B, providing the opportunity
to probe its physical conditions.  The determination of changes in the
radio transmission properties, including the dispersion and rotation
measures, will potentially allow the plasma density and magnetic field
structure to be probed.  Additionally, the $\sim$70-yr period
geodetic precession will cause the line-of-sight to sample different
trajectories through the magnetosphere of B. To a lesser extent, the
21-yr orbital precession will also cause changes in the trajectory.

Close inspection of the flux density of the pulses from A (Fig.~5)
reveals that a short occultation occurs, centered upon its superior
conjunction (when the line-of-sight passes only 0.15~lt-s from B).
The duration of the occultation is about 20--30~sec.  Because the
relative transverse velocity is about 660~km/s, the eclipsing region
has a lateral extent of $\sim$15,000~km or $\sim$0.05~lt-s, about 10\%
of the light cylinder radius of B. As far as we can tell at present,
the eclipse occurs in every orbit and its extent is essentially the
same at 680~MHz and 1390~MHz, limiting the possible
interpretations. One clue may come from the fact that the rate of
spin-down energy loss from A is $\sim$3600 times greater than that from
B. In fact, at the light cylinder radius of B, the energy density of
the relativistic wind from A is about two orders of magnitude greater
than that from B, ensuring that the wind from A will penetrate deep
into B's magnetosphere. We find that the energy densities due
to the spin-down luminosity (assumed isotropic) emitted by A and the
magnetic field of B are in balance at a distance $\sim$0.2~lt-s from
B, about 40\% of the light cylinder radius of B, assuming that the
magnetic field can be calculated using the standard dipole formula
\cite{mt77}. Within this distance, the energy field of B will
dominate, while A will dominate outside. Even though this picture is
roughly consistent with the lateral extent of the observed eclipse of
A, it is likely that the strong departure of B's magnetosphere from an
ideal case makes such calculations uncertain.

Because the point of pressure-balance is deep within the magnetosphere
of B, the actual penetration of the wind from A into B's magnetosphere
will be a function of the orientation of the rotation and/or magnetic
axes of B relative to the direction of the wind and hence will depend
on the precessional and orbital phases of B. This is the most likely
explanation for the large flux density and pulse-shape changes of B
which are seen to vary with orbital phase (Fig.~1). We note that these
changes are essentially the same between 680~MHz and 3030~MHz.  It
seems most likely to us that such a broadband modulation arises from
the impact of energy in the form of particles, gamma-rays or 44-Hz
electromagnetic radiation from the millisecond pulsar A upon the
magnetosphere of B. We note that there is a strong indication of
significant unpulsed radio emission from the system: the time-averaged
pulsed flux density of the pulsars is about 1.8~mJy at 1390~MHz
(Table~1), compared with a total flux density at this frequency of
7~mJy \cite{bdp+03}.  The $\sim$5~mJy unpulsed emission probably
arises in the impact region described above.  We find it remarkable,
with much of the magnetosphere of B blown away by the wind of A, that
B still works as a pulsar. This suggests that the radio emission is
probably generated close to the neutron star, providing a direct
constraint on the emission height.

\paragraph*{Conclusion.}
We have detected the binary companion of the millisecond pulsar
J0737$-$3039 as a pulsar, making this the first known double-pulsar
system. This discovery confirms the neutron star nature of the
companions to recycled pulsars in eccentric binary systems and
validates the suggested evolutionary sequences in which a companion
star, having spun up the pulsar, forms a young pulsar in a supernova
explosion \cite{sv82}.  The highly-relativistic nature of this compact
system opens up opportunities for much more stringent tests of
relativistic gravitation than have been possible previously.  Not only
have we already measured four quantities attributable to, and
consistent with, general relativity, but the mass-ratio $R$ is a new
high-precision constraint that is independent of gravitational
theories.  Within a year or so we expect to measure the orbital decay
due to emission of gravitational radiation. If the intrinsic value due
to gravitational-wave damping can be extracted, it will allow tests of
radiative aspects of gravitational theories mixed with strong-field
effects.  On somewhat longer time-scales of a few years we expect to
detect several other relativistic effects such as geodetic precession
of the pulsars' spin axes, spin-orbit coupling and other
deviations, making this a superb testbed for relativity.

The detection of the companion as a pulsar also opens up the
possibility of using each pulsar to probe the magnetosphere of the
other.  The energy flux from the millisecond pulsar is strongly
affecting the pulse emission process in the companion and eclipses of
the millisecond pulsar by the companion are also seen. Future
measurements of orbital variations in pulse shapes, amplitudes,
polarization and timing over a range of radio frequencies will give
fascinating insights into magnetospheric processes in pulsars.

\bibliography{journals,modrefs,psrrefs,crossrefs,notes}

\begin{thebibliography}{10}
\bibitem{rhealth}
 This is correct up to so-called first post-Newtonian ($(v/c)^2$) order
  \cite{dt92} and any intersection of the PK-parameter lines must be located on
  the $R$-line which will only deviate from that shown in Fig.~3 by corrections
  of order $(v/c)^{4}$.
\bibitem{massdef}
 This is correct due to the usage of the DD timing model in our analysis and
  its definition of mass \cite{dd85,dd86}.
\end{thebibliography}

\bibliographystyle{Science}


\begin{scilastnote}
\item We would like to thank John Sarkissian and other members of the
Parkes multibeam team for their kind help with making the observations
described in this paper. Extensive use was made of the PSRCHIVE pulsar
analysis system developed by Aidan Hotan and colleagues (see
http://astronomy.swin.edu.au/pulsar).  We are grateful to Norbert Wex
and Gerhard Sch{\"a}fer for useful discussions. The Parkes radio
telescope is part of the Australia Telescope which is funded by the
Commonwealth of Australia for operation as a National Facility managed
by CSIRO.  MB, AP and ND'A acknowledge financial support from
the Italian Ministry of University and Research (MIUR) under the
national program {\em Cofin 2001}.  FC is supported by NSF, NASA,
and NRAO. DRL is a University Research Fellow funded by the Royal
Society.
\end{scilastnote}


\begin{table}
\caption{Observed and derived parameters of PSRs~J0737$-$3039A and B
using the DD timing model \cite{dd85,dd86}.  Standard
($1\,\sigma$) errors are given in parentheses after the values and are
in units of the least significant digit(s). The parameters $A$, $B$
and $\delta_r$ in the DD model were assumed to be zero in the
analysis. The distance is estimated from the dispersion measure and a
model for the interstellar free electron distribution \cite{tc93}.}
\begin{center}
\begin{tabular}{lcc}
\hline
 & & \\ 
Pulsar & PSR~J0737$-$3039A & PSR~J0737$-$3039B \\
Pulse period $P$ (ms) & 22.69937855615(6) & 2773.4607474(4) \\
Period derivative $\dot{P}$ & $1.74(5) \times 10^{-18}$ & $0.88(13)\times 10^{-15}$ \\
Epoch of period (MJD) & 52870.0 & 52870.0 \\
Right ascension (J2000) & $07^{\rm{h}}37^{\rm{m}}51^{\rm{s}}.247(2)$ & $-$ \\
Declination (J2000) & $-30^\circ 39' 40''.74(3)$ & $-$ \\
Dispersion measure DM (cm$^{-3}$pc) & 48.914(2) & 48.7(2)\\
Orbital period $P_{\rm{b}}$ (day) & 0.102251563(1) & $-$ \\
Eccentricity $e$ & 0.087779(5) & $-$ \\
Epoch of periastron $T_0$ (MJD) & 52870.0120589(6) & $-$ \\
Longitude of periastron $\omega$ (deg) & 73.805(3) & 73.805 + 180.0 \\
Projected semi-major axis $x=a \sin i/c$ (sec) & 1.41504(2) & 1.513(4) \\
Advance of periastron $\dot{\omega}$ (deg/yr) & 16.90(1) & $-$ \\
Gravitational redshift parameter $\gamma$ (ms) & 0.38(5) &  \\
Shapiro delay parameter $s$ & $0.9995(-32,+4)$ & \\
Shapiro delay parameter $r$ ($\mu$s) & $5.6(-12,+18)$ &  \\
RMS timing residual ($\mu$s) & 27 & 2660 \\
Flux density at 1390 MHz (mJy) & 1.6(3) & 0--1.3(3) \\
 & & \\
Characteristic age $\tau$ (My) & 210 & 50 \\
Surface magnetic field strength $B$ (Gauss) & $6.3\times 10^9$ & $1.6\times 10^{12}$ \\
Spin-down luminosity $\dot E$ (erg/s) & $5800\times10^{30}$ & $1.6\times10^{30}$ \\
Mass function (M$_\odot$) & 0.29097(1) & 0.356(3) \\
Distance (kpc) & \multicolumn{2}{c}{$\sim$0.6} \\
Total system mass $m_{\rm{A}}+m_{\rm{B}}$ (M$_\odot$) & \multicolumn{2}{c}{2.588(3) } \\
Mass ratio $R \equiv m_{\rm{A}}/m_{\rm{B}}$ & \multicolumn{2}{c}{1.069(6) } \\
Orbital inclination from Shapiro $s$ (deg) & \multicolumn{2}{c}{87(3)} \\
Orbital inclination from $(R,\dot\omega)$ (deg) & \multicolumn{2}{c}{$87.7(-29,+17)$} \\
Stellar mass from $(R,\dot\omega)$ (M$_\odot$) & 1.337(5) & 1.250(5) \\
 & & \\
\hline
\end{tabular}
\end{center}
\end{table}

\clearpage

\noindent {\bf Fig.~1.} The intensity of radiation of B as a function
of orbital longitude relative to the ascending node of its orbit.
Each panel shows a grey-plot of intensity over a phase range of 0.1 of
pulsar rotational period, centered on the pulsed emission.  Panels
(A), (B) and (C) present observations at 680~MHz, 1390~MHz and
3030~MHz respectively (summed over 6 orbits at 1390~MHz and 2 orbits
at each of the other frequencies), showing the similarity of the
intensity variation over the wide frequency range and the changing
pulse-shape with orbital longitude.  Longitude in this diagram (and in
Figs.~2 and 5) is strictly the sum of the longitude of periastron and
the true anomaly.  The vertical line at longitude 270$^\circ$
represents the longitude of inferior conjunction of B, when the two
stars are at their closest on the sky and B is the nearer to the
Earth.

\noindent {\bf Fig.~2.} The effect of the Shapiro delay caused by the
gravitational potential of B seen in the timing residuals of A.
Residuals were averaged into 75 equal bins of orbital longitude.  Top:
timing residuals obtained by subtracting the model defined in Table~1
from the observed TOAs.  Bottom: as top display, but with the Shapiro
delay parameters $r$ and $s$ set to zero.

\noindent {\bf Fig.~3.} The observational constraints upon the masses
$m_{\rm{A}}$ and $m_{\rm{B}}$.  The colored regions are those which
are excluded by the Keplerian mass functions of the two pulsars.
Further constraints are shown as pairs of lines enclosing permitted
regions as predicted by general relativity (see text for details): (a)
the measurement of the advance of periastron $\dot{\omega}$, giving
the total mass $m_{\rm{A}}+m_{\rm{B}}=2.588\pm0.003$ M$_{\odot}$
(dashed line); (b) the measurement of
$R \equiv m_{\rm{A}}/m_{\rm{B}}=x_{\rm{B}}/x_{\rm{A}}=1.069\pm0.006$ (solid
line); (c) the measurement of the gravitational redshift/time dilation
parameter $\gamma$ (dot-dash line); (d) the measurement of Shapiro
parameter $r$ giving $m_{\rm{B}}=1.2\pm0.3$ M$_{\odot}$
(dot-dot-dot-dash line) and (e) Shapiro parameter $s$ (dotted line).
Inset is an enlarged view of the small square which encompasses the
intersection of the three tightest constraints, with the scales
increased by a factor of 16. The permitted regions are those between
the pairs of parallel lines and we see that an area exists which is
compatible with all constraints, delineated by the solid blue region.

\noindent {\bf Fig.~4.} The physical configuration of the binary
system, at conjunction, as on 19 August 2003 (MJD~52870), showing the
relative sizes of the two orbits and B's magnetosphere.  At
conjunction, the two neutron stars are separated by $\sim$2.8~lt-s or
about 800,000~km.  Top: view from above the orbital plane with the
Earth to the right.  The shaded segments indicate orbital phases where
B is detected strongly (see Fig.~1).  Also shown is the apsidal line,
which is the major axis of the orbits, and the line-of-nodes, which is
the intersection of the orbital plane and the plane through the center
of mass of the system which is normal to the line-of-sight.  Bottom:
view from the side, showing the passage of the line-of-sight from A to
the Earth through the magnetosphere of B.  The approximate position of
the pressure balance between the relativistic wind from A and the
magnetic field of B is indicated.

\noindent {\bf Fig.~5.} The variation in flux density of A (in
arbitrary units) at 680~MHz and 1390~MHz, around superior conjunction
(i.e. longitude 90$^\circ$).  The data are presented with 5-sec time
resolution and show the eclipse of the pulsar by the magnetosphere of
B.

\clearpage

\begin{figure}
\caption{}
\centerline{\psfig{file=fig1.ps,width=15cm}}
\end{figure}


\begin{figure}
\caption{}
\centerline{\psfig{file=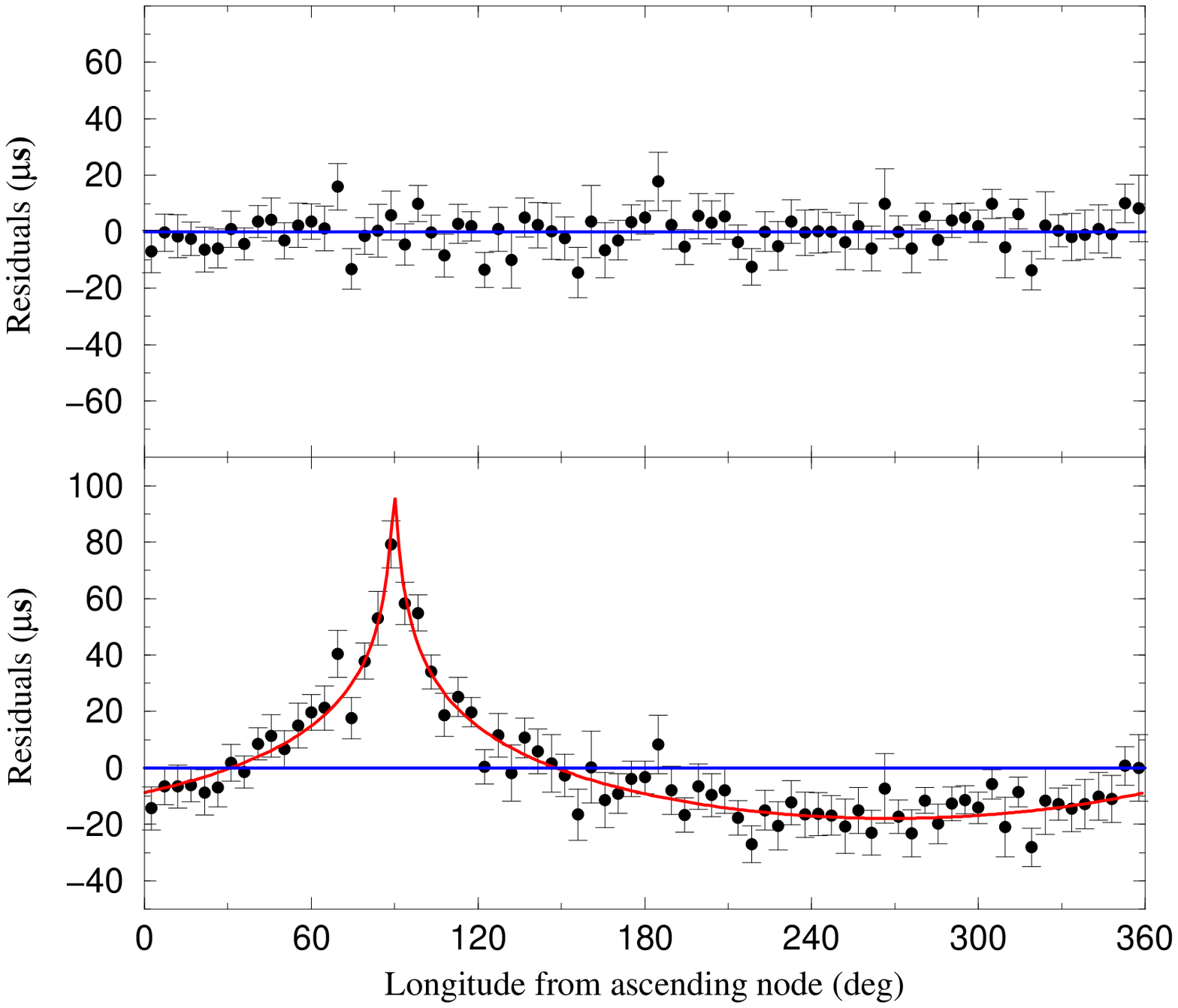,width=15cm}}
\end{figure}

\begin{figure}
\caption{}
\centerline{\psfig{file=fig3.ps,angle=-90,width=15cm}}
\end{figure}

\begin{figure}
\caption{}
\centerline{\psfig{file=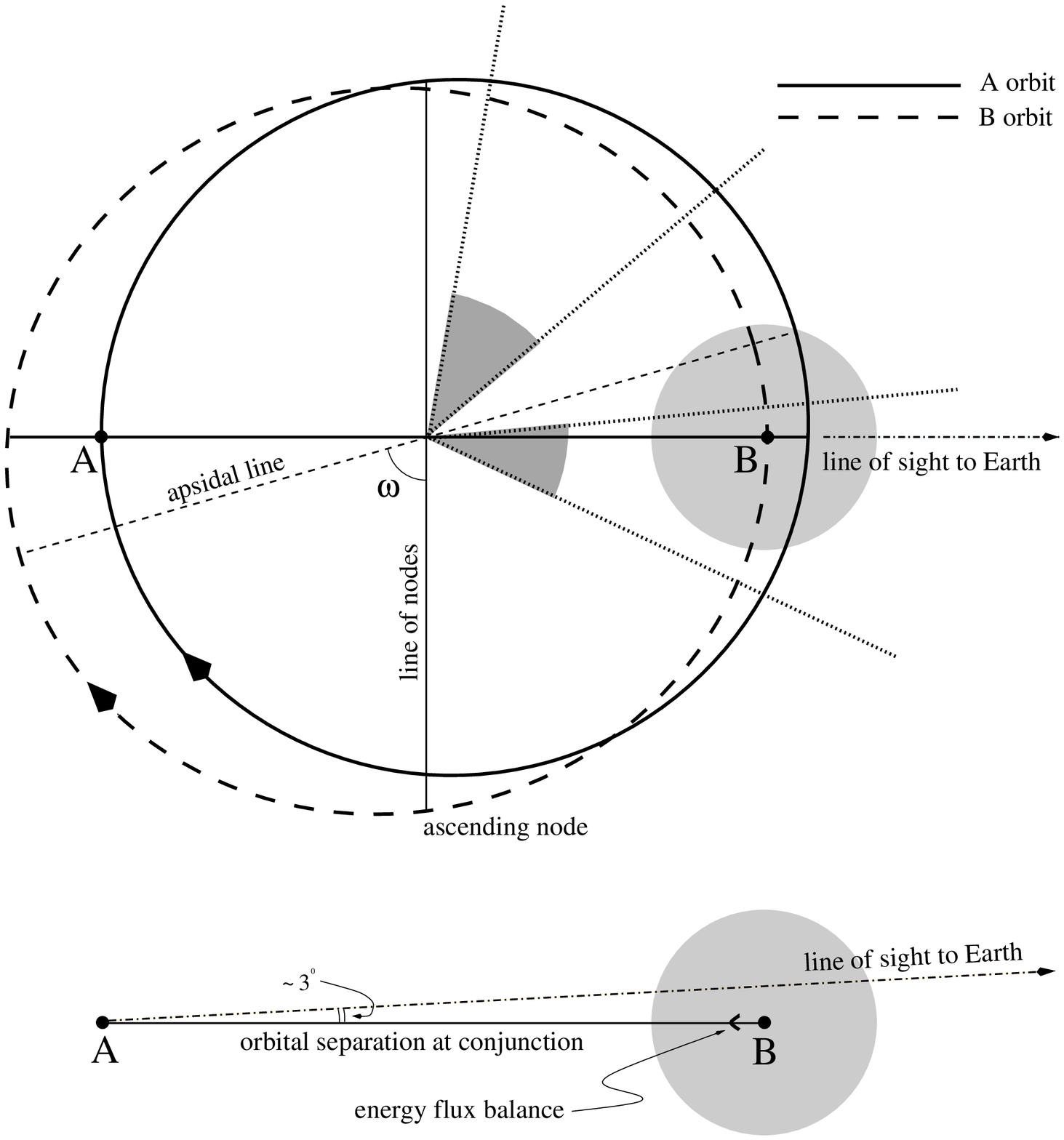,angle=0,width=15cm}}
\end{figure}

\begin{figure}
\caption{}
\centerline{\psfig{file=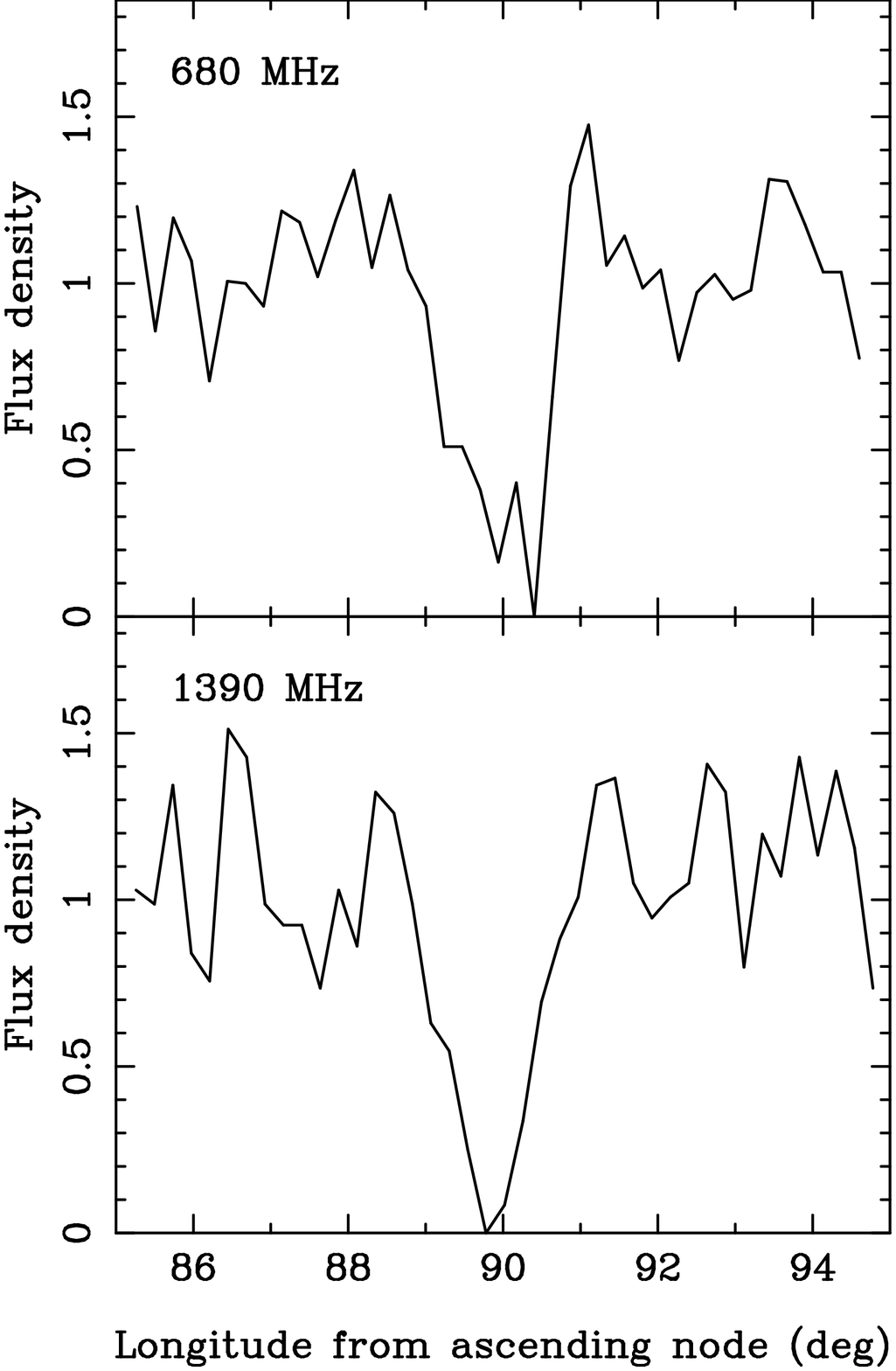,angle=0,width=15cm}}
\end{figure}

\end{document}